\documentclass{aa}

\usepackage{color}

\usepackage{graphicx}
\usepackage{txfonts}
\usepackage{natbib}

\bibpunct{(}{)}{;}{a}{}{,} % to follow the A&A style
\begin{document}

%\shorttitle{Long Gamma Ray Bursts from binaries}
%\shortauthors{TBD}

\title{Long Gamma Ray Bursts from binary black holes}
\author{Agnieszka Janiuk\inst{1}, Szymon Charzy{\'n}ski\inst{2}
\and
Micha{\l} Bejger\inst{3}
}

\institute{Center for Theoretical Physics,
Polish Academy of Sciences, Al. Lotnikow 32/46, 02-668 Warsaw, Poland \\
\email{agnes@cft.edu.pl}
\and
Faculty of Mathematics and Natural Sciences,  Cardinal Stefan
Wyszy\'nski University,  ul. W\'oycickiego 1/3, 01-938 Warsaw, Poland \\
\email{szycha@cft.edu.pl}
\and
Copernicus Astronomical Center, Bartycka 18, 00-716 Warsaw,
Poland \\
\email{bejger@camk.edu.pl}
}

   \date{Received ...; accepted ...}
  \abstract
{}
{
We consider a scenario for the longest duration gamma ray bursts,
resulting from the collapse of a massive rotating star in a close binary
system with a companion black hole.
}
   {
The primary black hole born during the core collapse
 is first being spun up and increases its mass during
the fallback of the stellar envelope just after its birth.
As the companion black hole enters the outer envelope, it provides
an additional angular momentum to the gas.
After the infall and spiral-in towards
the primary, the two black holes merge inside the circumbinary disk.
}
{
The
second episode
of mass accretion and high final spin of the post-merger black hole
prolongs the gamma ray burst central engine activity. The observed events
should have two distinct peaks in the electromagnetic signal, separated by
the gravitational wave emission. The gravitational recoil of the burst
engine is also possible.
}
{}
\keywords{black hole physics; accretion; gamma ray bursts}
  \authorrunning{A. Janiuk et al.}
  \titlerunning{GRBs from black hole binaries}
   \maketitle

%%%%%%%%%%%%%%%%%%%%%%%%%%%%%%%%%%%%%%%%%%%%%%%%%%%%%%%%%%%%%%%%%%%%%%%%
\section{Introduction}
%%%%%%%%%%%%%%%%%%%%%%%%%%%%%%%%%%%%%%%%%%%%%%%%%%%%%%%%%%%%%%%%%%%%%%%%
Gamma ray bursts are transient sources of extreme brightness
observed on the sky with isotropic distribution. Their prompt phase
lasts between a fraction of a second and few hundreds seconds, and the
long duration events are believed to originate from collapsing
massive stars. In the collapsar model, a newly born black hole (BH)
surrounded by a transient disk accreting a part of the fall-back stellar
envelope helps launching relativistic jets
\citep{1993ApJ...405..273W, 1999ApJ...524..262M}
. These polar jets give rise to the
gamma rays, produced far away from the 'engine' in the circumstellar
region
(see e.g., the reviews by \citealt{2004IJMPA..19.2385Z},
\citealt{2004RvMP...76.1143P}).
%Zhang \& Meszaros 2004; Piran 2005; Metzger 2010).
The model is supported by observed associations of many gamma ray bursts
with bright supernovae
%(e.g. Woosley \& Bloom 2006).
\citep{2006ARA&A..44..507W}.
These are
the brightest I b/c type explosions of the so-called 'hypernov\ae' which
constitute about 10\% of this class \citep{2007astro.ph..2338F}.
%(Fryer et al. 2007).
What seems to be
most important for a pre-supernova star to become a GRB progenitor, is its
high rate of differential rotation \citep{2004ApJ...607L..17P}.
%(Podsiadlowski et al. 2004).
The massive stars that are progenitors of GRBs, at the end of their
lives go through the stage of a Wolf-Rayet star \citep{2007ARA&A..45..177C}.
%(Crowther 2007).
Such a star may be spun up by the interaction in a binary system. In
addition, the loss of angular momentum through the stellar wind may be
avoided when
the metallicity of the star is sufficiently low (\citet{2005A&A...443..643Y},
%Yoon \& Langer 2005,
see also \citet{2010MNRAS.405...57S}).

A possible configuration in the binary star evolution history would be a
close binary that consists of a massive OB star and a compact remnant resulting
from an earlier core collapse. Such a system, i.e., a high mass X-ray binary
\citep{WellsteinL1999}, will evolve to form a WR star--BH binary, such
as the well known Cyg X-3 system discovered in our Galaxy \citep{Kerkwijk1992},
or the extragalactic sources IC 10 X-1 and NGC 300 X-1
\citep{BauerB2004,Carpano2007}. Here, we consider the final stage of evolution
of such a binary, in which the BH ultimately enters the massive star's
envelope and spins it up (essentially, a common envelope phase of the binary).
This process triggers the collapse of the core, possibly via the tidal
squeezing interaction \citep{LuminetM1985}, and may provide an additional source of power to the
GRB event. Similar scenarios were proposed in the past, for
example by \citet{ZhangF2001,BarkovK2010}, in which three phases can be
distinguished: the spiral-in of the BH inside the envelope, possibly
with a spherical accretion of some surrounding gas and transfer of orbital
angular momentum into the envelope; increase of the accretion rate through the
high-angular momentum shells of matter onto the BH residing already, or
newly born, in the center (see \citealt{Chevalier2012} for a description of a
collapse event triggered by the inspiral of the compact object to the central
core of the companion star); and final accretion of the remaining gas during
the ultimate GRB explosion accompanied by the jet ejection.  In
addition to the electromagnetic signal, such a process will also be followed by
a characteristic gravitational-wave signal due to the collapse of the core
into the BH, but mostly because of the binary BH inspiral and
merger.
%As the system is embedded in matter, a modeling similar to the
%simulations of massive BH binaries in galaxy mergers may be applicable
%\citep{farris12}.

This article is composed as follows: Sect.~\ref{sect:models} describe the
models used to estimate the basic parameters of the process.
Sect.~\ref{sect:bhtorus} describes a model of the BH surrounded by a torus,
Sect.~\ref{sect:homo} describes the homologous accretion without a mass loss,
whereas Sect.~\ref{sect:torloss} considers the case of strong winds during the
accretion of the in-falling shells. Sect.~\ref{sect:premergsum} gathers
the pre-merger scenarios. The model of the binary BH merger is described in
Sect.~\ref{sect:bhmerger}. Sect.~\ref{sect:discconc} contains discussion
and conclusions.

Through the text, the subscript 1 denotes the primary BH, a result of
the collapse of the primary component in the binary system. Subscript 2
denotes the companion BH, whereas subscript 3 marks the final BH, which
results from a merger of 1 and 2 black holes.

%%%%%%%%%%%%%%%%%%%%%%%%%%%%%%%%%%%%%%%%%%%%%%%%%%%%%%%%%%%%%%%%%%%%%%%%
\section{Collapsing star with a companion black hole}
\label{sect:models}
%%%%%%%%%%%%%%%%%%%%%%%%%%%%%%%%%%%%%%%%%%%%%%%%%%%%%%%%%%%%%%%%%%%%%%%%
\subsection{Model of a BH surrounded by a torus}
\label{sect:bhtorus}
%%%%%%%%%%%%%%%%%%%%%%%%%%%%%%%%%%%%%%%%%%%%%%%%%%%%%%%%%%%%%%%%%%%%%%%%

First, we test the predictions of the model of the collapsing star that
encounters a companion BH. We use a simple ``toy model'' calculation
to quantify the behaviour of the rotating BH in the center of the
collapsing star and an accreting torus embedded in its envelope. We focus on
the evolution of the BH spin and changes in the accretion rate within
the torus, to make predictions on the duration and power available for the gamma ray
burst.

The primary is a collapsing star, whose iron core has just formed a central
BH of mass $M_{1,init}$.  The distribution of density in the envelope
is the same as used in \citet{janiuk08}, and is taken from the spherically
symmetric pre-supernova star of a mass of 25 $M_{\odot}$ \citep{Woosley95}.
The size of the star is $R_{\rm out} \approx 6 \times 10^{13}$ cm, so the
free-fall timescale from this radius is about $t_{\rm ff} \approx 10^{7}$ s.
The rotation of the stellar envelope leads to formation of the torus, i.e.,
high angular momentum shells located in the equatorial plane, that will
subsequently accrete onto the core.
%to account for
%the polar jets presumably produced via the coupling with accretion onto the
%spinning black
%hole via magnetic fields, and ultimately responsible for the observed
%gamma ray burst emission.
The specific angular momentum distribution is in general given by
\begin{equation}
l_{\rm spec} = l_{0} f(\theta) g(r),
\end{equation}
where the normalization is scaled to the critical angular
momentum, $l_{0}/l_{\rm crit}=x$. We express $l_{\rm crit}$ as
\begin{equation}
l_{\rm crit} = {2 G M_{1} \over c} \sqrt{2 - a_{1} + 2 \sqrt{1 - a_{1}}},
\end{equation}
where $a_{1}$ is the primary black hole dimensionless spin parameter.
%, a ratio of the total BH angular momentum $J_{1}$
%to its mass $M_{1}$;
The above equation gives the condition for the formation of
a disk with the angular momentum exceeding that of the marginally bound orbit
\citep{bardeen72}. The dependence on polar angle $\theta$ is
\begin{equation}
f(\theta) = 1 - |{\cos \theta}|.
\end{equation}
In the present model we neglect the radial dependence, i.e., we take
into account differential rotation only (see however \citet{janprog08} for the
discussion of other rotation laws).

Both mass and spin of the primary BH change during the collapse of the
envelope, as the massive shells accrete onto the center. The BH
absorbs only the angular momentum of the gas, which is smaller than the
critical one. The rotating torus, however, is supported by the gas which
angular momentum is larger than the critical. The value of $l_{\rm crit}$
changes during the collapse, affecting the evolution of the BH spin and
conditions for the torus existence.

We then introduce a secondary (companion) BH
of a mass $M_{\rm 2}$ and negligible spin, falling
into the envelope of the primary star at the onset of its collapse.
As the companion BH moves from the radius $r$ to $r-\Delta r$ inside the envelope,
it transfers its specific orbital angular momentum to the shells:
\begin{equation}
\Delta l = {d J_{\rm 2} \over dM} = {d J_{\rm 2} \over dr} / {dM \over dr} \approx
{M_{\rm 2} \over 2} \sqrt{ {G r \over M(r)}} (1+ \ln {r_{\rm 2} \over r}),
\end{equation}
where $M(r)$ is the mass of the envelope inside the radius $r$.
We assume here that the companion BH orbital angular momentum is Keplerian,
$J_{\rm 2} = M_{\rm 2} \sqrt{G M(r) r}$ (see \citealt{BarkovK2010}).
In addition, we assume explicitly that the companion enters the envelope close
to the equatorial plane,
so that the
specific angular momentum is transferred as
$l_{\rm spec} = l_{\rm spec} + \Delta l f(\theta)$.
%using the $f(\theta)$ prescription.}
 %\tb{so that $f(\theta)\simeq 1$, and hence $l_{\rm spec}\simeq l_{0}$.}

%%%%%%%%%%%%%%%%%%%%%%%%%%%%%%%%%%%%%%%%%%%%%%%%%%%%%%%%%%%%%%%%%%%%%%%%
\subsection{Homologous accretion without mass loss}
\label{sect:homo}
%%%%%%%%%%%%%%%%%%%%%%%%%%%%%%%%%%%%%%%%%%%%%%%%%%%%%%%%%%%%%%%%%%%%%%%%
First, we analyze the simplest scenario, where the whole envelope collapses via
homologous accretion of subsequent shells. Therefore not only
the high angular momentum, but also the low $l_{\rm spec}$ gas contributes
to the black hole evolution.

\begin{figure}
\includegraphics[width=\columnwidth]{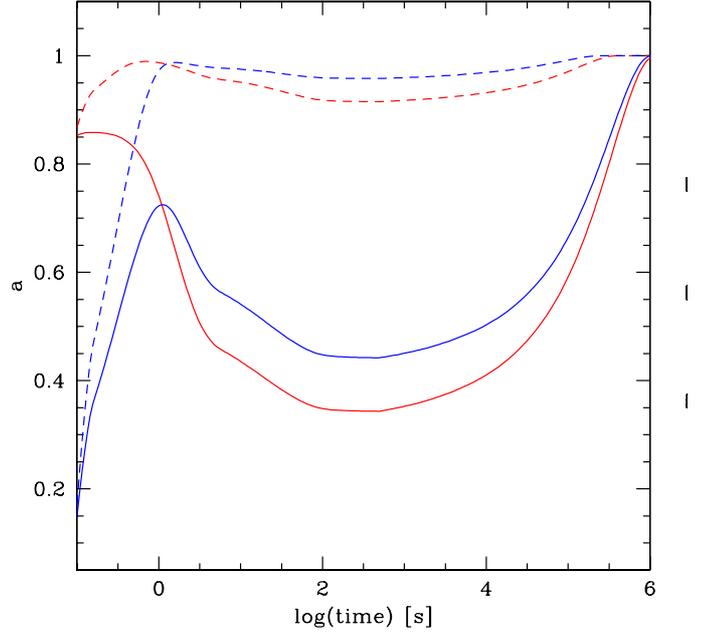}
\caption{Evolution of the primary BH spin during the collapse of the stellar
envelope.
Blue lines are for $a_{\rm 1,init}=0.1$,
%green lines for $A_{\rm init} =0.5$ and
and red lines for $a_{\rm 1,init}=0.85$.
The solid and dashed lines show the models with the envelope's angular
momentum normalized with $x=l_{\rm spec}/l_{\rm crit}=1.5$ and 7, respectively.
The model assumes homologous accretion of the envelope shells onto the BH,
and no wind. The envelope is spun up by the companion BH of a mass
$M_{2} = 3 M_{\odot}$.
The time is given as the free-fall timescale, so it scales with $M_{1}$.
}
\label{fig:akerr}
\end{figure}

Fig.~\ref{fig:akerr} shows the evolution of the BH spin with time.
The time is calculated as a free-fall time of the envelope shell that
surrounds the central BH of mass $M_{\rm BH} (\rm t)$.
We show two examples of the specific angular momenta of the envelope, parameterized by
$x=l_{\rm spec}/l_{\rm crit}=1.5$ (solid lines) and $x=7.0$ (dashed lines).
% {NO $l_{0} = l_{\rm spec}$?}

%-----------------------------------------------------------------------
%-----------------------------------------------------------------------
Initially, the BH spin grows in a short timescale due to accretion of high
angular momentum material from the rotating torus.  After it decreases,
sometimes even below the initial value. This is due to accretion of the low
angular momentum gas, i.e., with $l_{\rm spec}$ smaller than the critical value
$l_{\rm crit}$ for a current BH mass and spin \citep{janiuk08}. If the spin-up
of the envelope is neglected, the rotationally-supported torus is present only
temporarily. Here however, the companion spins up the envelope again, so that
the high angular momentum gas is available to spin up the primary BH.
Finally, the furthest shells of the envelope collapse onto the center.
The duration of this episode is not sensitive to the initial BH spin and
angular momentum normalization. The latter affects the minimum spin of the
primary BH during this episode, and in our examples it is about $a_{\rm 1,min}=0.4-0.5$ for
$x=1.5$, and about $a_{\rm 1,min}=0.9-0.95$ for $x=7.0$.
In any case, at the end of the
collapse the primary BH will spin at almost maximum rate.

Our calculations show two accretion episodes. The first lasts
%between a fraction of a second
up to a few hundreds of seconds, depending on angular momentum
in the envelope, $x$, and the primary BH spin $a_{\rm 1, init}$.
The accretion rate in the torus is initially
almost 0.1 $M_{\odot}$s$^{-1}$ but steeply
decreases, following the density profile
of gas in the subsequent shells of the envelope.
The second accretion episode begins at $\sim 500$ seconds and is governed
by the presence of the companion. This episode lasts until the whole envelope
has collapsed; the accretion rate (estimated simply
as $\Delta m_{\rm torus} / \Delta t$)
 now rises to above
1 $M_{\odot}$s$^{-1}$, because of larger mass available in the shells.

%%%%%%%%%%%%%%%%%%%%%%%%%%%%%%%%%%%%%%%%%%%%%%%%%%%%%%%%%%%%%%%%%%%%%%%%
\subsection{Torus accretion with mass loss}
\label{sect:torloss}
%%%%%%%%%%%%%%%%%%%%%%%%%%%%%%%%%%%%%%%%%%%%%%%%%%%%%%%%%%%%%%%%%%%%%%%%
In the above section, the accretion rate was estimated using the free-fall
timescale. Primary BH was spun up to a maximum rotation, practically
regardless of its initial spin. Moreover, the homologous
accretion of shells led to effective increase of the BH mass and at the end
of the simulation it simply equals to the initial mass of the pre-collapse star.

Now, we consider a more realistic scenario, when accretion onto the primary
BH proceeds through a thick, viscous disk and a substantial fraction of
the envelope mass is not accreted but lost to the massive winds.
Such winds were discussed in \citep{1999ApJ...524..262M}
and have also been found in numerical
simulations by \citet{2006MNRAS.368.1561M},
who reported on their mildly relativistic velocities and intermediate opening
angles.
In our recent work \citep{Janiuk2013, 2012IJMPS...8..352J}, we also
studied the
 neutrino cooled disk/winds in the GRB central engine, via the
magnetohydrodynamical simulations of accretion.
%The detailed results
%of our extensive GR MHD simulations will be presented in a separate article,
We found that
 the mass taken away by the winds and not accreted onto the black hole
through the event horizon might reach the fraction of even 50-72\%.
This fraction depends on the parameters such as black hole mass and spin,
and might also be sensitive to the adopted initial distribution of the
specific angular momentum in the gas. These winds are launched by the
magnetic pressure and are bright in neutrinos that cool the central engine.
We checked that for some
models the winds appear to be bound, so that they would actually results in
some large scale circularization movements, however for other models
the wind velocity exceeds the escape velocity.
The mass loss of 72\% must therefore be treated
as an upper limit.

Here,
%we simply adopt this ratio to be a free parameter, and
we assume a fiducial value of a maximum fraction of the wind mass loss, so
that the mass accreted onto
the primary BH is 28\% of the shell,
the remaining fraction being lost from the system.

The viscous timescale in the accretion disk is given by
\begin{equation}
t_{\rm visc} = 250 \left(\frac{\alpha \delta^{2}}{0.01}\right)^{-1}
\left({r \over 10^{3} r_{g}}\right) \left({M_{\rm 1} \over 10 M_{\odot}}\right)\ {\rm s},
\end{equation}
where $\alpha$ is the viscosity parameter, and $M_{\rm BH}$ and
$r_{\rm g}$ are the BH mass and gravitational radius, respectively.
The ratio of disk thickness to radius $h/r \equiv \delta$ can be estimated
from the model of a neutrino cooled disk in the GRB central engine
(Janiuk \& Yuan 2010); it is about $0.1-0.3$ for the neutrino
transparent (low accretion rate) models.

\begin{figure}
\includegraphics[width=\columnwidth]{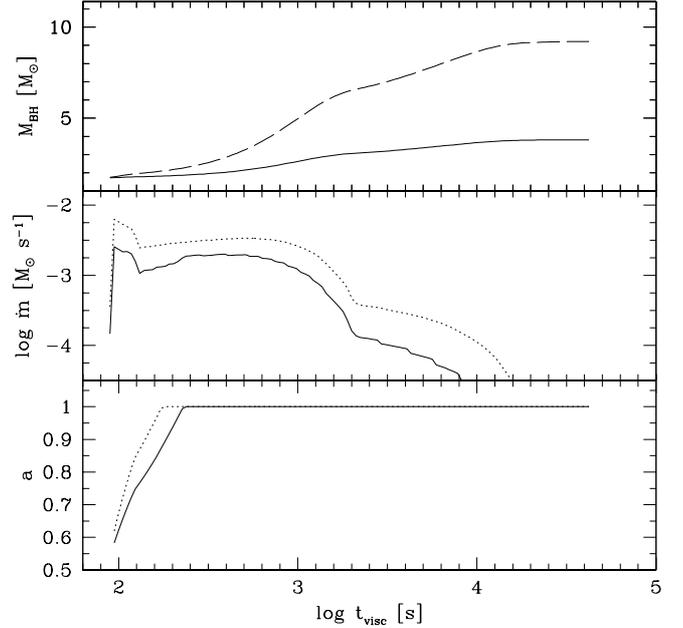}
\caption{Mass of the primary BH (top), accretion rate (middle) and the
primary BH spin (bottom panel) as a function of time during the accretion of
the viscous torus. In the top panel, the model assumes the specific angular
momentum in the envelope: $x=1.5$ to compare two cases: either 28\% of the
torus mass (solid line) and the rest is blown out, or its total mass (dashed
line) is fully accreted. In the middle and bottom panels, the
models assume the wind outflow, and the two lines show the cases
with different angular momentum: $x=1.5$ (solid) and $x=7.0$ (dotted).
}
\label{fig:3panels}
\end{figure}

In the Figure \ref{fig:3panels} we plot the BH mass, accretion rate and
BH spin as a function of time. The time is expressed in viscous time
scales of the accreting torus at the distance $r$ from the center. The final
BH spin in this model is always maximal and is reached very quickly
after the onset of accretion.
In this scenario, most of the envelope's
mass is blown out with the wind and not accreted onto BH, while some of the
matter is accreted but it contributes to the primary black hole
spin up more than to its mass increase.
We assume that 72\% of the envelopes' mass was ejected with the
massive wind outflow.
%-----------------------------------------------------------------------
%-----------------------------------------------------------------------
The mass of the central BH grows as long as the torus exists. The torus is
supported by both the specific angular momentum in the envelope and by the
companion. For the smallest $l_{\rm spec}$ normalizations we have tested,
$x=1.5$, and no companion, we obtained $M_{\rm 1,final}=4.4 M_{\odot}$ at the end of
the simulation (the result is for the particular pre-collapse star density
distribution and depends on the assumed fraction of the torus mass taken out by
the wind). If no wind outflow was assumed, and the whole torus  mass was
accreted onto the BH, then its final mass was $M_{\rm 1,final}=8.4M_{\odot}$.
This
value is very weakly dependent on the angular momentum distributions in
the stellar envelope.

We calculate the instantaneous accretion rate (middle panel of Fig.~\ref{fig:3panels})
in the torus as the ratio between the mass of an accreting shell
and the local viscous timescale, $\dot m = \delta m_{\rm torus}/\delta t_{\rm visc}$.
Initially, the accretion rate is peaking at about
0.01 $M_{\odot}$ s$^{-1}$, for a large specific angular momentum in the envelope.
The second peak in the accretion rate lasts much longer, but the accretion rate
is less than during the first peak, because of a smaller density of the accreted
material. Finally, the accretion rate drops below $10^{-4} M_{\odot}$s$^{-1}$
even though the torus persists, because of smaller density and long viscous timescale.

\begin{figure}
\includegraphics[width=\columnwidth]{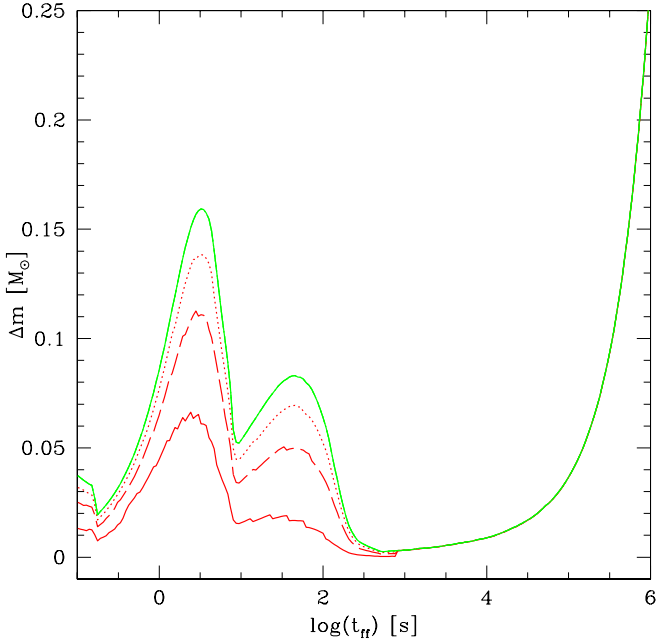}
\caption{Mass of the envelope shell as a function of the free-fall time.  Green
solid line is the total mass in the shell, while the red lines correspond with
the mass contained in the rotationally supported torus. Here, the solid, long dashed and
short dashed lines show the models with the specific angular momentum
normalized with $x=1.5$, 3 and 7, respectively. All models assume that 28\% of
the stellar envelope is accreted onto the BH, and the rest of the mass
is evacuated as a massive wind. The outer envelope is additionally spun up by
the companion BH, which after the accretion of the inner shells is merging
with the primary, and subsequently will accrete the outer shells.}
\label{fig:dmwind}
\end{figure}

The accretion in the torus proceeds through three episodes, as shown in
Fig.~\ref{fig:dmwind}. The Figure shows the mass of a given accreting shell
versus the free-fall timescale at the initial distance of this shell.  The
shells that are closest to the center, have their free falling timescales below
$\sim 900$ seconds, and masses up to 0.15 $M_{\odot}$.  The exact value of the
shell mass which is contained within the rotationally supported torus, is
sensitive to the magnitude of specific angular momentum (the Figure shows three
different lines for $x=1.5$, 3.0 and 7.0). The total shell mass (i.e. torus
plus polar regions) is of course the same for any $x$, nevertheless it is still
governed by the onion-like distribution of the elements in the envelope. This
leads to the two distinct peaks.
%-----------------------------------------------------------------------
%-----------------------------------------------------------------------
The outer shells are made of lighter elements and are less dense; their
mass increases only due to the larger volume. Because the outermost envelope is
first spun up by the companion, the effect of the
intrinsic angular momentum distribution is not important and basically
all the outer envelope contributes to the rotationally supported torus.
Therefore the third accretion episode is the fall-back of the material from
the outer shells, spun up by the companion and rotating in the torus.

%%%%%%%%%%%%%%%%%%%%%%%%%%%%%%%%%%%%%%%%%%%%%%%%%%%%%%%%%%%%%%%%%%%%%%%%
\subsection{Summary of the pre-merging scenarios}
\label{sect:premergsum}
%%%%%%%%%%%%%%%%%%%%%%%%%%%%%%%%%%%%%%%%%%%%%%%%%%%%%%%%%%%%%%%%%%%%%%%%

We considered the scenarios of a homologous or torus accretion onto the newly
formed BH in the core of the primary star. We tested a range of angular
momentum normalizations in the envelope. We also included the possibility of
the wind taking away most of the mass from the rotationally supported torus.
In this scenario, the BH mass grows more slowly, however its spin is
still very quickly growing to the maximum limit, regardless of the initial
value of the stellar core rotation.

%The most representative are two scenarios. First,

Due to the accretion of the inner shells of the star's envelope
onto the core, the primary BH mass increases to
$M_{\rm 1,final}$ and the spin obtains the value of $a_{\rm 1,final}$.
These values are checked at the time when the companion black hole
approaches the primary core to the shells which have already been accreted
(i.e., $r_{2}(t_{\rm final}) = r^{\rm k}$ and $k$ is the number of the currently
accreting shell). In our model, this moment
corresponds to the distance of $r_{2}=10^{11}$ cm and the free fall
timescale at this distance is $t_{\rm final}\sim 530$ s
(scaling with mass $M_{1}$).

In the homologous accretion scenario, the primary black hole mass is at this moment equal to
about $M_{\rm 1,final}= 9 M_{\odot}$ (for our assumed progenitor star model,
but independently on the initial core rotation rate and specific
angular momentum distribution in the envelope).
%{\bf depending on parameters ??}.
If the initial spin was $a_{\rm 1,init}=0.5$, it temporarily dropped
due to the in-fall of low angular momentum material, and then increased.
The final value of the spin, for a moderately rapid rotation in the envelope,
given by our parameter $x=3$, was about $a_{\rm 1,final}=0.69$.
%\tb{(is this value correct or should be close to 1?)}.

This primary BH will then merge with the companion and we assume its mass of
$M_{2} = 3 M_\odot$ and negligible spin.
After the merger, the remaining mass of the envelope, which in this example
will be equal to
about $M_{\rm env}(t_{\rm final})\sim 16 M_{\odot}$,
will accrete onto the product of the merger.

The second scenario is the accretion through the viscous, rotationally
supported torus onto the primary BH, under the assumption that most of the
material is blown out with a massive wind. We assume that only 28\% of mass
accretes onto the core and contributes to its growing mass and spin.  The
resulting BH will nevertheless be spinning at the maximum rate, as all the
accreting material has large specific angular momentum.  The mass of the
primary BH after this accretion episode is about
$M_{\rm 1,final} = 3.8\ M_\odot$.  As the
companion BH mass is again assumed $3\ M_\odot$, the final BH is produced of a
merger of two comparable mass BHs.
The product $M_{3}$ subsequently accretes the remaining envelope.
The mass of the gas available for accretion in the final episode is in this example equal to about $6 M_{\odot}$.

In both scenarios outlined above, the mass of the final BH, $M_{3}$,
 and the remaining torus mass
are comparable. We do not follow here
this final accretion process numerically, as it would require an enormous
computational power to run a full GR MHD simulation in a non-stationary metric.
 In the stationary
Kerr metric, such simulations of accretion onto a single black hole
have recently been shown elsewhere (e.g.
 \citet{2012MNRAS.423.3083M}).
We aimed however to treat in more detail the binary black hole merger
process, which timescale is much shorter than the timescale of
accretion of the distant torus,
and may be treated separately from the surrounding matter, i.e.
in the vacuum approximation.
Below we present our several numerical simulations,
focusing in particular on the two distinct scenarios that led to
different initial parameters of the merging black holes.

%%%%%%%%%%%%%%%%%%%%%%%%%%%%%%%%%%%%%%%%%%%%%%%%%%%%%%%%%%%%%%%%%%%%%%%%
\section{Binary black hole merger}
\label{sect:bhmerger}
%%%%%%%%%%%%%%%%%%%%%%%%%%%%%%%%%%%%%%%%%%%%%%%%%%%%%%%%%%%%%%%%%%%%%%%%
%%%%%%%%%%%%%%%%%%%%%%%%%%%%%%%%%%%%%%%%%%%%%%%%%%%%%%%%%%%%%%%%%%%%%%%%
\subsection{Physics of the model}
\label{sect:mergphysics}
%%%%%%%%%%%%%%%%%%%%%%%%%%%%%%%%%%%%%%%%%%%%%%%%%%%%%%%%%%%%%%%%%%%%%%%%
The simulation covers the very last stage of the evolution of binary BH
system, when the separation of the components becomes so small, that
the phases of inspiral, merger and ringdown can be tracked. This is
the stage of the evolution for which the full set of Einstein
equations needs to be solved numerically to model the geometry of
spacetime in order to obtain reasonable results.

The initial state of the system under consideration consists of two
black holes in quasi circular orbits with mass ratio varying from 2 to 3. The
more massive black hole carries also spin perpendicular to the orbital plane,
the second component is spinless. The direction of the
initial spin vector of the black hole coincides with the direction of the
orbital angular momentum of the binary system.

We  performed several runs of simulations, for different values
of spin of the rotating black hole. The parameters for each run are
presented in Table \ref{tw}. The initial separation of components is
the same for each run and is equal to $6 M$, where the value of $M$
is close to the $ADM$ mass of the whole system
%The ADM stands for Arnowitt, Deser and Misner formalism
\citep{PhysRev.116.1322}, defined as
the mass measured by a distant observer in an asymptotically flat
space time.

Despite the fact that we do not change mass parameters of punctures
representing black holes (for numerical details see next section)
for each run, the ADM mass of the system varies, since we vary the
spin which contributes to the total ADM mass. This is a known
property of rotating black holes, for example for analytical Kerr
solution we have:
\begin{equation}
M_{\rm ADM}=\sqrt{M_{\rm irr}^2+\frac{S^2}{4M_{\rm irr}^2}}
\end{equation}
where $M_{\rm ADM}$ is the  ADM mass, $S$ is the spin of black hole and
$M_{\rm irr}$ is the irreducible mass - namely the mass related to the
area of the event horizon (the mass of nonrotating black hole with
the same area of event horizon,
% the spin can be ,,extracted'' via the Penrose process,
for more details see \citet{gravitation}).

%The values of ADM mass of the system are given also in table
%\ref{tw}, and they are close to $1M$ for each run.

%%%%%%%%%%%%%%%%%%%%%%%%%%%%%%%%%%%%%%%%%%%%%%%%%%%%%%%%%%%%%%%%%%%%%%%%
\subsection{Numerics}
\label{sect:mergnumerics}
%%%%%%%%%%%%%%%%%%%%%%%%%%%%%%%%%%%%%%%%%%%%%%%%%%%%%%%%%%%%%%%%%%%%%%%%

We use the fifth release of the
%(code name \texttt{Lovelace}, released 28-05-2012)
%\cite{EinsteinToolkit:web}
\texttt{Einstein Toolkit}\footnote{http://einsteintoolkit.org}
\citep{Loffler:2011ay} based on \texttt{Cactus Computational Toolkit}
\citep{Goodale:2002a}.
%The \texttt{Einstein Toolkit} consists of many
%modules called \emph{thorns}. These modules are designed to realize
%specified tasks.

The initial data are provided by the \texttt{TwoPunctrures} thorn
\citep{Ansorg:2004ds}. This module solves numerically the binary
puncture equations for a pair black holes \citep{Brandt:1997tf}. The
initial state of space time is described by extrinsic curvature in
the Bowen-York form \citep{Bowen:1980yu}, for given mass, momentum
and spin of each puncture. These are the controlled parameters in
the first section of Table \ref{tw}.

The evolution is performed by the \texttt{McLachlan} module
\citep{Brown:2008sb} which is a numerical implementation of the $3+1$
split of Einstein equations, solving the Cauchy initial value problem using
the Baumgarte-Shapiro-Shibata-Nakamura (BSSN) method \citep{Shibata:1995we, Baumgarte:1998te,
Alcubierre:2000xu}.

The simulation is performed on the cartesian grid with the size of
$60\times60\times60 M$ and resolution of $dx=dy=dx=2 M$ (runs R1 - R7 in Table \ref{tw}),
or the size of $48\times48\times48 M $ and resolution
of $dx=dy=dz=1.6$ (runs R8 and R9). We use 7
levels of the adaptive mesh refinement in two regions around singularities.
Each
refinement is by the factor of 2 and the radii of refinement regions
are: $0.5$, $1$, $2$, $4$, $8$ and $16$ around each singularity. The
regions of refined grid follow the positions of singularities
%which are localized  by the thorn \texttt{PunctureTracker}.
%Regridding is performed by the thorn \texttt{Carpet}
\citep{Schnetter:2003rb}. We
assume that the space time has a reflection symmetry with respect to
the plane spanned by the initial momenta of components of the BH
binary system,
%This symmetry is implemented by the \texttt{ReflectionSymmetry} module,
which reduces the number of grid
points by the factor of two.
%The thorn \texttt{AHFinderDirect}
%\cite{Thornburg:2003sf} is used to localize
The apparent horizons are localized
around the components of the BH system and around final merged black
hole after it forms \citep{Thornburg:2003sf}. The proper integrals over the isolated horizons
are calculated
%by the thorn \texttt{QuasiLocalMeasures} \cite{Dreyer:2002mx}
to extract the values of mass and spin of the
merged black hole \citep{Dreyer:2002mx}.

In all simulations we note the effect of gravitational recoil of the final BH.
This effect is illustrated in Fig.~\ref{pic-trajectories}. This is known
effect, see for example \cite{tichy2007}.
The calculation of the exact value of
the recoil speed requires the evaluation of the momentum carried away by the
gravitational radiation during the merger. We did not analyze the
gravitational radiation in the first series of simulations performed. In order to estimate the recoil speed we have done last run of the simulation (R9 in the Table \ref{tw}) once again with analysis of radiation included. To calculate total momentum carried by radiation we have followed algorithm described by \cite{alcubierre}. We use the formula for $d\vec{P}/dt$ in terms of coefficients $A^{lm}$ of multipole expansion of the Weyl scalar $\psi_4$. The coefficients $A^{lm}$ are computed by the thorns \texttt{WeylScal4} and \texttt{Multipole} on the sphere of radius $22M$ and $l$ ranging from 2 to 4. We integrate $d\vec{P}/dt$ over the time of the simulation to get total linear momentum radiated from the system through gravitational waves. Since total momentum has to be conserved we are able to compute recoil of the merged black hole.

%%%%%%%%%%%%%%%%%%%%%%%%%%%%%%%%%%%%%%%%%%%%%%%%%%%%%%%%%%%%%%%%%%%%%%%%%
\subsection{Results}
\label{sect:results}
\begin{figure}
%\plotone{run_12.ps}
\begin{center}
\includegraphics[width=\columnwidth]{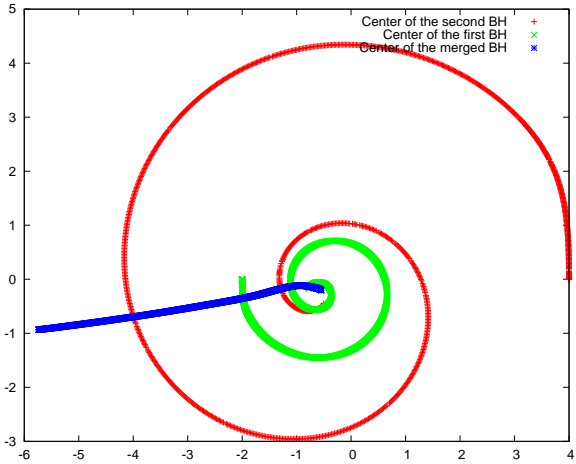}
\end{center}
\caption{Trajectories of the component BHs for an exemplary
 simulation described in
Table~\ref{tw} (run R9). Simulation covers the last two orbits of
BH binary (red and green) and the formation of the final BH (blue).}
\label{pic-trajectories}
\end{figure}

%%%%%%%%%%%%%%%%%%%%%%%%%%%%%%%%%%%%%%%%%%%%%%%%%%%%%%%%%%%%%%%%%%%%%%%%%
Results of the simulations are presented in Table~\ref{tw}. We also
present a plot of the BH trajectories for the  run R9
 ( Fig.~\ref{pic-trajectories}).
We note that there is a saturation of the final
BH spin at about $a_{3}\sim0.8$,
%which is reached for mergers of component BHs with $a_1=0.3$ (that is,
(primary BH has $a_1= 0 - 0.9$, and companion
BH $a_2=0$ in all runs). Models with $a_1=0.9$ do not result in
a significant growth of the final BH spin.

In the last five simulations (R6-R10) the initial momentum of the BH system was
varied.The initial spin parameter $a_1$ was kept constant, $a_1=0.9$
in runs R6-R9.
These initial data correspond therefore to orbits that are not necessarily
circular.  The values the final BH spin do not change
significantly with the initial orbital angular momentum of the system. This can
be understood as a manifestation of the emission of gravitational waves: part
of angular momentum is radiated away with the gravitational radiation. Above a
certain limit, the increase of the total angular momentum of the initial system
(orbital and spin) does not result in the increase of spin of the final BH; the
amount of angular momentum radiated away increases, however.

We have done quantitative analysis of gravitational radiation for the last two
runs of simulations (R9 and R10 in the Table \ref{tw}).
The direction of the recoil is irrelevant, since it  depends on which phase
of the last orbit the components of the system meet (in our simulations must
remain in the orbital plane, since the reflection symmetry is assumed).
The velocity of the final BH depends on spins and masses of the components
and basically it is on the order of a few thousands km/s (see, e.g., \cite{tichy2007}).  In this particular cases we obtained the values of recoil speeds to be approximately 200 km/s and 300 km/s, for the runs R9 and R10, respectively.

The runs that roughly correspond to the pre-merging scenarios outlined
in Section 2, are R5 or R9 for the first (i.e. homologous accretion) scenario
and R10 for the second (i.e. torus accretion and wind outflow) scenario.
The second one, being more realistic in the physical collapse models,
leads to the approximately equal mas ratio of the merging holes
 and high spin of the primary. Our merger simulations confirm therefore, that
a high recoil velocity is obtained in this case, albeit not as large
as in the runs with black hole
mass ratio of about 3 and moderate or high spins.

\begin{table*}
\begin{center}
\caption[]{Summary of the binary BH merger
 models. Parameters $m_1$, $p_1$ and $m_2$,
 $p_2$ are mass and $x$-component of momentum of the primary and the companion BH,
respectively. $s_1$ is the spin of the primary BH (the $z$-component).
 ADM values of the initial state are computed from the given controlled
 parameters: $M_1$ and $M_2$ are the ADM masses of the components,
 $M$ is the total ADM mass of the system, $a_1 = s_1/M_1^2$ is the
 dimensionless spin parameter of the first component. Final state
 $M_3$ and $a_3 = s_3/M_3^2$ are ADM mass and dimensionless spin parameter
 of the final BH.} \label{tw}
\begin{tabular}{|r|c|c|c|c|c||c|c|c|c|c||c|c|}
 \hline
 & \multicolumn{10}{|c||}{ Initial state} & \multicolumn{2}{|c|}{Final state}
\\
 \hline
 & \multicolumn{5}{|c||}{ Parameters}
 & \multicolumn{5}{|c||}{ Computed ADM values}
 & \multicolumn{2}{|c|}{ADM values}
\\
 \hline
 run & $m_1$ & $m_2$ & $p_1$ & $p_2$ & $s_1$ & $M_1$ & $M_2$ & ${M_1}/{M_2}$ & $M$ & $a_1$ & $M_3$ & $a_3$
\\
 \hline
 R1 & 0.632 & 0.316 & -0.121 & 0.121 & 0   & 0.652 & 0.337 & 1.93 & 0.976 & 0 & 0.961
 & 0.581 %08
\\
 \hline
 R2 & 0.632 & 0.316 & -0.121 & 0.121 & 0.1 & 0.666 & 0.338 & 1.97 & 0.989 & 0.226 & 0.972
 & 0.650 %10
\\
 \hline
 R3 & 0.632 & 0.316 & -0.121 & 0.121 & 0.3 & 0.749 & 0.339 & 2.21 & 1.070 & 0.535 & 1.051
 & 0.741 %11
\\
 \hline
 R4 & 0.632 & 0.316 & -0.121 & 0.121 & 0.5 & 0.853 & 0.342 & 2.49 & 1.172 & 0.687 & 1.157
 & 0.762 %12
\\
 \hline
 R5 & 0.632 & 0.316 & -0.121 & 0.121 & 0.7 & 0.958 & 0.346 & 2.77 & 1.273 & 0.764 & 1.261
 & 0.757 %13
\\
 \hline
 R6 & 0.632 & 0.316 & -0.121 & 0.121 & 0.9 & 1.057 & 0.35 & 3.02 & 1.368 & 0.806 & 1.358
 & 0.79 %14
\\
 \hline
 R7 & 0.632 & 0.316 & -0.135 & 0.135 & 0.9 & 1.054 & 0.349 & 3.02 & 1.373 & 0.81 & 1.354
 & 0.802 %15
\\
 \hline
 R8 & 0.632 & 0.316 & -0.16 & 0.16 & 0.9 & 1.052 & 0.349 & 3.01 & 1.382 & 0.813 & 1.342
 & 0.771 %18
\\
 \hline
 R9 & 0.632 & 0.316 & -0.172 & 0.172 & 0.9 & 1.052 & 0.35 & 3.0 & 1.387 & 0.813 & 1.344
 & 0.761 %19
\\
 \hline
 R10 & 0.54 & 0.445 & -0.138 & 0.138 & 0.3 & 0.6 & 0.445 & 1.35 & 1.031 & 0.788 & 0.982
 & 0.779 %mass_ratio_1_run_03
\\
% \hline
% 23 & 0.632 & 0.316 & 9 & -0.105 & 0.105 & 0.9 & 1.05 & 0.337 & 3.11 & 1.372 & 0.817 & 1.279
% & 0.99
%\\
 \hline
\end{tabular}
\end{center}
\end{table*}

%-----------------------------------------------------------------------
%-----------------------------------------------------------------------

%%%%%%%%%%%%%%%%%%%%%%%%%%%%%%%%%%%%%%%%%%%%%%%%%%%%%%%%%%%%%%%%%%%%%%%%
\section{Discussion and conclusions}
\label{sect:discconc}
%%%%%%%%%%%%%%%%%%%%%%%%%%%%%%%%%%%%%%%%%%%%%%%%%%%%%%%%%%%%%%%%%%%%%%%%
The scenarios presented here consider the long gamma ray burst case,
resulting from the massive rotating star collapse that occurs in a close binary
system with a companion BH. The event can be divided into three stages. First,
the innermost shells of the massive primary accrete onto the core that has
collapsed to a BH. These massive shells add the mass and angular momentum to
this newly-born BH, and the ultimate outcome depends on the fraction of
envelope that is blown away from the system through the wind. The companion BH
spins up the outer shells of the rotating envelope and subsequently falls into
the gap after the inner torus has been accreted.  The second stage consists
 of the
merger of two BHs, surrounded by a remnant circumbinary disk. The product of
the merger will have a net spin and a recoil velocity that depends on the
binary parameters.  In the third stage the final BH accretes the remaining
material that constitutes the outer rotationally-supported torus.

We studied two classes of models: with and without the massive wind launched
from the primary star. In the first scenario, the primary BH is spun up to the
maximum rotation rate due to accretion of only the high angular momentum
material, while it accretes a moderate amount of mass only. Our calculations
show, that the mass of the BH increases from 1.7 up to 3.8 $M_{\odot}$. It then
merges with the companion BH of mass $M_{2} = 3\ M_\odot$. In the second
scenario, the primary BH accretes both high and low angular momentum material
from the envelope in the first stage. Therefore its mass prior to the binary BH
merger is large, e.g., 9.2 $M_{\odot}$. The spin however may not increase
significantly and even drop below the starting value. The details depend on the
magnitude of specific angular momentum deposited in the stellar envelope. For
our testing parameters, we obtained $a=0.40$, $0.69$ and $0.94$ for the $l_{\rm
spec}$ normalizations $x=1.5$, 3.0 and 7.0, respectively.

We found no significant dependence of our results on the companion mass,
provided it is a stellar mass BH on the order of 1-3 $M_{\odot}$.
In all the cases, we neglected the spin of the companion BH, as
well as the increase of its mass in the Bondi-like accretion during its
passage through the primary's envelope.
Prior to the merger of the binary BH system, in the central engine we have
either a maximally spinning and moderately massive, or more massive and
moderately spinning BH and circumbinary torus. After the merger, the final BH
mass is between $M_{3} = 7 - 12 M_{\odot}$ and the mass of the remnant torus
is about 16 $M_{\odot}$. We may expect that most of this material
will eventually be lost from the envelope through the massive outflows.
The transfer of the orbital angular momentum from the companion to the envelope
leads to the longer lifetime of the rotationally supported torus around the
primary BH. The accretion rate through this torus is small, however the
infalling matter contributes to increasing BH mass and spin.

In the present work, we consider a fiducial value of the mass loss parameter
from the accreting torus due to the powerful winds.
% Narayan \& Kumar (2008, MNRAS 388, 1729)
\citet{2008MNRAS.388.1729K}
discuss the problem of mass fallback in the long GRB central
engine and notice that the advection dominated part
of the accretion flow generates a strong mass outflow.
In their model, the 14 Solar mass star ends up with the black hole of 10
solar masses, which means that 4 solar masses (28 per cent) was lost
from the system through the wind.

%Lindner et al. (2010, ApJ, 713, 800)
\citet{2010ApJ...713..800L}
in their hydrodynamical
simulations of collapsar use also the 14 Solar
mass Wolf-Rayet star with low metallicity (of 0.01 $Z_{\odot}$), which results
in a very small mass loss.
These authors use the radius
dependent angular momentum profile $l(r) sin^{2}(\theta)$ with
the magnitude at
3/4 mass radius of $8\times 10^{17}$ cm$^{2}$ s$^{-1}$.
In the models discussed in \citet{Janiuk2013},
on which the fiducial value of the wind parameter was assumed here,
the specific angular momentum was about $6 \times 10^{16} - 2\times 10^{17}$
cm$^{2}$s$^{-1}$ and therefore  the centrifugal force could
help driving the wind outflow. However, as we included also the magnetic
fields and neutrino emission in those calculations, the
 more powerful winds were launched.
Still, in not all models the winds appeared to be gravitationally unbound,
therefore this fiducial number we assumed in the present paper
must treated as an upper limit for the wind mass loss. Nevertheless, we note
that the results presented in Section 2 are mostly sensitive to
the adopted accretion scenario (i.e. the homologous vs. torus accretion) and
the assumed wind fraction in case of torus accretion
 does not affect them in a great detail.

As for the progenitor star and its mass loss rate due to the wind, the rate
is uncertain and observationally poorly constrained.
%Heger et al. (2003)
\citet{2003ApJ...591..288H}
found that a low metallicity reduces mass loss
in supernovae (see also
%Woosley and Heger 2006).
\citet{2006ApJ...637..914W}).
%Mapelli et al (2013)
\citet{2013MNRAS.429.2298M}
use the power-law dependence of mass loss rate
$\dot M~Z^{\mu}$, where $\mu=0.5-0.9$ is the index for the main sequence stars.
However, in case of Luminous Blue Variable stars and Wolf Rayet stars
this scaling might be different
%(Vink and de Koter 2005).
\citep{2005A&A...442..587V}.
%Dwarkadas (2013, MNRAS, 434, 3368)
\citet{2013MNRAS.434.3368D}
study the supernova remnants
and find that the mass loss and wind velocities in the
Wolf-Rayet stars are on the order of $10^{-5} M_{\odot}$ yr$^{-1}$ and 2000
km s$^{-1}$, respectively.
In case of red supergiants, the mass loss rate is larger, but the
wind velocity is smaller.

Observationally, at least some long GRBs that are associated with supernovae,
must have strong winds which make them not 'failed' supernovae. These are e.g.
GRB021211
%(Della Valle et al. 2003, A\&A, 406, L33).
\citep{2003A&A...406L..33D}.
On the other hand, GRB 060614, 100 seconds long duration, had no supernova signatures
%(Dell Valle et al. 2006, Nature, 444, 1050).
\citep{2006Natur.444.1050D}.

The merger of two black holes occurs when the inner torus has completely
accreted onto the primary and a clean gap formed at the distance of approximately
twice the orbital separation (see, e.g., \citealt{Shi:2012,Farris:2012}).
For stellar mass BHs the timescale of the merger is of the order of milliseconds.
The merger event in our simulations should occur at about $\sim 1700 - 2000$,
which is related to the timescale
of the progenitor star collapse and the crossing time of the secondary
component through the outer envelope shells.
%seconds after the onset of the primary collapse.
Accretion after the merger
proceeds then onto the product BH, with a smaller accretion rate than in the
first episode.
This is because despite a large mass of the outer envelope shells,
most of it is blown out and not accreted, while
the viscous timescale at large radii is long.
The total duration of this phase is determined by the size and
mass of the primary star. In our model, the parameters of the star
taken for a simulation imply the viscous timescale
at $t_{\rm visc}(R_{\rm out}) \approx 10^{7}$ s. This is therefore the
source of a resulting GRB afterglow emission, observable at lower energies
for the following few months after the prompt event.
% is higher and

The total observed event should have two distinct
components in the electromagnetic
signal, separated by a gravitational wave emission.
One of the accompanying effects will also be the product black hole recoiled due to
the gravitational waves.
%{\bf The total event lasts for ...??? seconds and is aimed to explained the longest duration GRBs ???
%Bogdanovic et al. (2007)
This effect has been intensively discussed recently for the scenarios of
the supermassive black hole mergers. For instance, \citet{bogdanovic2007}
proposed a scenario of gas-rich binary black hole
mergers. In this scenario, torques from gas accretion align the spins of
two black holes and their orbital axis with the large-scale disks.
The authors argue, that this alignment prevents large kicks from
the gravitational radiation recoil and helps explain the observations
that ubiquity of black holes remain in the galaxy cores, despite their past
mergers.

This reasoning is based on the results of the simulations.
%Baker et al (2006),
For instance, \citet{2008PhRvD..78d4046B}
modeled
the coalescence of non-spinning black holes with different mass ratios.
%and  Hermann et al (2007) modeled the BHs with spin axis parallel
%and anti-parralell to orbital axis -  only in preprints...
Also,
%Gonzales et al (2007)
\citet{2007PhRvL..98w1101G}
and
\citet{2007PhRvL..98w1102C}
%Campanelli et al (2007)
studied the cases with general spin orientations.
These simulations show that for mergers with BHs of
low spins or the spins aligned, the maximum speeds of the kick are below 200 km/s.
For the spins oppositely directed and large $a$ values, the kicks exceed 4000 km/s,
while the escape speed from the galaxy core is below 1000 km/s \citep{2004ApJ...607L...9M}.
The simulations' results are not conclusive to say however, that large kicks
are prevented by the coaligned spins of merging black holes. For instance,
\citet{tichy2007} performed simulations of equal mass BHs
with spins of $\sim 0.8$ and random orientations. They showed that all
recoil velocities are large, between 1000 and 2000 km/s.
Higher spins lead to even larger kicks. The kick velocity depends also
on the mass ratio. Above-mentioned authors did not calculate this, but they expect that
smaller kicks will be obtained for unequal mass case.
This is consistent with analytical
estimates presented in \citet{bogdanovic2007}.
In our simulations, the mass ratios and spins are approximately
either (i)
q=0.3, $a_{1} = 0.9$ and $a_{2}=0$ or (ii) q=1.35, $a_{1}=0.8$ and $a_{2}=0$,
so the ratio between the two kicks will be about 0.8 with this
simplified formula.
Therefore the kick of 200 km/s obtained in the first case (unequal masses)
would scale up to about 250 km/s for the second case.
We obtained only a kick of 300 km/s in model R10, which
seems to be in a rough agreement with these analytical estimates,
taking also into account some numerical uncertainties.

The escape velocities estimated for a sample of short GRBs' host
galaxies from \citet{2010MNRAS.405...57S} with median mass of
$M_{\rm host} = 1.3 10^{9} M_{\odot}$ and the 80\% light radius of $r_{80}= 3.3$ kpc would be about 2280 km s$^{-1}$. It is therefore not possible that
the merger product simulated in our models will leave the
host galaxy. Such extreme kick velocities could however be obtained
if both merger components had very large spins.

The gravitational wave emission is estimated at $\lesssim 10\%$ of the rest
mass-energy of the system; similar figures can be obtained using the analytic
phenomenological formulae derived from numerical-relativity simulations by
\citet{Barausse:2012}.  Our simulation covers the merger phase only, and the
resulting gravitational wave emission is of the order of a few per cent (see
Table~\ref{tw} for the ADM mass differences).

By analogy to our current understanding of a two supermasive BHs system
residing in the centers of merging galaxies, we suspect that the interaction
with the gas-rich environment will facilitate the transfer of the kinetic
energy and orbital angular momentum from the binary system to the
gas\footnote{The observations of binary galaxies in the process of merging are
at present still poorly sampled. The only source for which the orbital modeling
finds a tight BH system, i.e., a sub-parsec solution is OJ287
\citep{2012MNRAS.427...77V}.  The other pairs of supermassive black holes are
rather wide, with separations on the order of hundreds of parsec
\citep{2011ApJ...736..125K}, and the indirect evidence for the presence of
binaries comes mostly from their semi-periodic lightcurves or the observations
of 'wiggles' in the radio jets, e.g., \citet{2009ApJ...705L..20X}.
Therefore the observational tests of such scenarios are also limited.}. This
should result in ''speeding up'' the inspiral and fewer orbits before the
merger. However, at smaller distances the two BHs may be orbiting in a region
relatively cleared of matter, surrounded by a circumbinary disk/torus and the
accretion from it may temporarily increase the binary system angular momentum,
possibly prolonging the inspiral phase.  The problem of exact GW signature from
such system depends on many unknown factors, like the recoil received during
the creation of a second BH and subsequent eccentricity of the orbit, as well
as the details of the MHD interaction with matter, and deserves separate
studies.

The electromagnetic emission
can be divided into three phases. First one is related to the collapse of the
progenitor star and creation of the primary BH, and its electromagnetic
emission is of the order of the SN emission. Second stage consists of the tidal
interaction of the binary BH system with the circumbinary accretion disk.
Rescaling the exploratory work results of \citet{Farris:2011} shows that for the
binary of $\simeq 10\ M_\odot$ the luminosity is $\simeq 10^{25}\ {\rm erg/s}$, much
fainter than the third and final phase, which pertains to the collapsar
scenario. In the case of substantial recoil, the final BH will drag the inner
part of the disk out of the system, and the electromagnetic counterpart will
be altered.

The progenitors of such systems are evolved binaries
in star-forming regions, most likely similar to Cyg X-1
or Cyg X-3. The first system most likely contains a
high mass black hole, while the latter shows significant contribution of
stellar wind component. The compact star in Cyg X-3
might already be a small mass
black hole
\citep{2013MNRAS.429L.104Z} or a neutron star that will eventually
collapse to a black hole during the inspiral phase.
%\citep{belczyn2006, portegies2000}.

\section*{Acknowledgments}
We thank Miko\l aj Korzy\'nski, Stefanie Komossa and Magda Kunert-Bajraszewska
 for helpful discussions.
This work was supported in part by the grants NN 203 512 638,
DEC-2012/05/E/ST9/03914 and 2011/01/B/ST9/04838 from the
Polish National Science Center.

\bibliographystyle{aa} % style aa.bst
%\bibliography{grb_bbh.bib} % your references Yourfile.bib
\bibliography{grb_bbh} % your references Yourfile.bib

\begin{thebibliography}{64}
\expandafter\ifx\csname natexlab\endcsname\relax\def\natexlab#1{#1}\fi

\bibitem[{Alcubierre(2008)}]{alcubierre}
Alcubierre, M. 2008, Introduction to 3+1 Numerical Relativity (Oxford
  University Press)

\bibitem[{Alcubierre {et~al.}(2000)Alcubierre, Br{\"u}gmann, Dramlitsch, Font,
  Papadopoulos, Seidel, Stergioulas, \& Takahashi}]{Alcubierre:2000xu}
Alcubierre, M., Br{\"u}gmann, B., Dramlitsch, T., {et~al.} 2000, Phys. Rev. D,
  62, 044034

\bibitem[{Ansorg {et~al.}(2004)Ansorg, Br{\"u}gmann, \& Tichy}]{Ansorg:2004ds}
Ansorg, M., Br{\"u}gmann, B., \& Tichy, W. 2004, Phys. Rev. D, 70, 064011

\bibitem[{Arnowitt {et~al.}(1959)Arnowitt, Deser, \& Misner}]{PhysRev.116.1322}
Arnowitt, R., Deser, S., \& Misner, C.~W. 1959, Phys. Rev., 116, 1322

\bibitem[{{Baker} {et~al.}(2008){Baker}, {Boggs}, {Centrella}, {Kelly},
  {McWilliams}, \& {van Meter}}]{2008PhRvD..78d4046B}
{Baker}, J.~G., {Boggs}, W.~D., {Centrella}, J., {et~al.} 2008, \prd, 78,
  044046

\bibitem[{{Barausse} {et~al.}(2012){Barausse}, {Morozova}, \&
  {Rezzolla}}]{Barausse:2012}
{Barausse}, E., {Morozova}, V., \& {Rezzolla}, L. 2012, \apj, 758, 63

\bibitem[{{Bardeen} {et~al.}(1972){Bardeen}, {Press}, \&
  {Teukolsky}}]{bardeen72}
{Bardeen}, J.~M., {Press}, W.~H., \& {Teukolsky}, S.~A. 1972, \apj, 178, 347

\bibitem[{{Barkov} \& {Komissarov}(2010)}]{BarkovK2010}
{Barkov}, M.~V. \& {Komissarov}, S.~S. 2010, \mnras, 401, 1644

\bibitem[{{Bauer} \& {Brandt}(2004)}]{BauerB2004}
{Bauer}, F.~E. \& {Brandt}, W.~N. 2004, \apjl, 601, L67

\bibitem[{Baumgarte \& Shapiro(1999)}]{Baumgarte:1998te}
Baumgarte, T.~W. \& Shapiro, S.~L. 1999, Phys. Rev. D, 59, 024007

\bibitem[{{Bogdanovi{\'c}} {et~al.}(2007){Bogdanovi{\'c}}, {Reynolds}, \&
  {Miller}}]{bogdanovic2007}
{Bogdanovi{\'c}}, T., {Reynolds}, C.~S., \& {Miller}, M.~C. 2007, \apjl, 661,
  L147

\bibitem[{Bowen \& York(1980)}]{Bowen:1980yu}
Bowen, J.~M. \& York, James~W., J. 1980, Phys. Rev. D, 21, 2047

\bibitem[{Brandt \& Br{\"u}gmann(1997)}]{Brandt:1997tf}
Brandt, S. \& Br{\"u}gmann, B. 1997, Phys. Rev. Lett., 78, 3606

\bibitem[{Brown {et~al.}(2009)Brown, Diener, Sarbach, Schnetter, \&
  Tiglio}]{Brown:2008sb}
Brown, J.~D., Diener, P., Sarbach, O., Schnetter, E., \& Tiglio, M. 2009, Phys.
  Rev. D, 79, 044023

\bibitem[{{Campanelli} {et~al.}(2007){Campanelli}, {Lousto}, {Zlochower}, \&
  {Merritt}}]{2007PhRvL..98w1102C}
{Campanelli}, M., {Lousto}, C.~O., {Zlochower}, Y., \& {Merritt}, D. 2007,
  Physical Review Letters, 98, 231102

\bibitem[{{Carpano} {et~al.}(2007){Carpano}, {Pollock}, {Wilms}, {Ehle}, \&
  {Schirmer}}]{Carpano2007}
{Carpano}, S., {Pollock}, A.~M.~T., {Wilms}, J., {Ehle}, M., \& {Schirmer}, M.
  2007, \aap, 461, L9

\bibitem[{{Chevalier}(2012)}]{Chevalier2012}
{Chevalier}, R.~A. 2012, \apjl, 752, L2

\bibitem[{{Crowther}(2007)}]{2007ARA&A..45..177C}
{Crowther}, P.~A. 2007, \araa, 45, 177

\bibitem[{{Della Valle} {et~al.}(2006){Della Valle}, {Chincarini}, {Panagia},
  {Tagliaferri}, {Malesani}, {Testa}, {Fugazza}, {Campana}, {Covino},
  {Mangano}, {Antonelli}, {D'Avanzo}, {Hurley}, {Mirabel}, {Pellizza},
  {Piranomonte}, \& {Stella}}]{2006Natur.444.1050D}
{Della Valle}, M., {Chincarini}, G., {Panagia}, N., {et~al.} 2006, \nat, 444,
  1050

\bibitem[{{Della Valle} {et~al.}(2003){Della Valle}, {Malesani}, {Benetti},
  {Testa}, {Hamuy}, {Antonelli}, {Chincarini}, {Cocozza}, {Covino}, {D'Avanzo},
  {Fugazza}, {Ghisellini}, {Gilmozzi}, {Lazzati}, {Mason}, {Mazzali}, \&
  {Stella}}]{2003A&A...406L..33D}
{Della Valle}, M., {Malesani}, D., {Benetti}, S., {et~al.} 2003, \aap, 406, L33

\bibitem[{Dreyer {et~al.}(2003)Dreyer, Krishnan, Shoemaker, \&
  Schnetter}]{Dreyer:2002mx}
Dreyer, O., Krishnan, B., Shoemaker, D., \& Schnetter, E. 2003, Phys. Rev. D,
  67, 024018

\bibitem[{{Dwarkadas}(2013)}]{2013MNRAS.434.3368D}
{Dwarkadas}, V.~V. 2013, \mnras, 434, 3368

\bibitem[{{Farris} {et~al.}(2011{\natexlab{a}}){Farris}, {Liu}, \&
  {Shapiro}}]{Farris:2012}
{Farris}, B.~D., {Liu}, Y.~T., \& {Shapiro}, S.~L. 2011{\natexlab{a}}, \prd,
  84, 024024

\bibitem[{{Farris} {et~al.}(2011{\natexlab{b}}){Farris}, {Liu}, \&
  {Shapiro}}]{Farris:2011}
{Farris}, B.~D., {Liu}, Y.~T., \& {Shapiro}, S.~L. 2011{\natexlab{b}}, \prd,
  84, 024024

\bibitem[{{Fryer} {et~al.}(2007){Fryer}, {Mazzali}, {Prochaska}, {Cappellaro},
  {Panaitescu}, {Berger}, {van Putten}, {van den Heuvel}, {Young},
  {Hungerford}, {Rockefeller}, {Yoon}, {Podsiadlowski}, {Nomoto}, {Chevalier},
  {Schmidt}, \& {Kulkarni}}]{2007astro.ph..2338F}
{Fryer}, C.~L., {Mazzali}, P.~A., {Prochaska}, J., {et~al.} 2007, ArXiv
  Astrophysics e-prints

\bibitem[{{Gonz{\'a}lez} {et~al.}(2007){Gonz{\'a}lez}, {Hannam}, {Sperhake},
  {Br{\"u}gmann}, \& {Husa}}]{2007PhRvL..98w1101G}
{Gonz{\'a}lez}, J.~A., {Hannam}, M., {Sperhake}, U., {Br{\"u}gmann}, B., \&
  {Husa}, S. 2007, Physical Review Letters, 98, 231101

\bibitem[{Goodale {et~al.}(2003)Goodale, Allen, Lanfermann, Mass{\'o}, Radke,
  Seidel, \& Shalf}]{Goodale:2002a}
Goodale, T., Allen, G., Lanfermann, G., {et~al.} 2003, in Vector and Parallel
  Processing -- VECPAR'2002, 5th International Conference, Lecture Notes in
  Computer Science (Berlin: Springer)

\bibitem[{{Heger} {et~al.}(2003){Heger}, {Fryer}, {Woosley}, {Langer}, \&
  {Hartmann}}]{2003ApJ...591..288H}
{Heger}, A., {Fryer}, C.~L., {Woosley}, S.~E., {Langer}, N., \& {Hartmann},
  D.~H. 2003, \apj, 591, 288

\bibitem[{{Janiuk} {et~al.}(2013){Janiuk}, {Mioduszewski}, \&
  {Moscibrodzka}}]{Janiuk2013}
{Janiuk}, A., {Mioduszewski}, P., \& {Moscibrodzka}, M. 2013, \apj, 776, 105

\bibitem[{{Janiuk} {et~al.}(2008){Janiuk}, {Moderski}, \& {Proga}}]{janiuk08}
{Janiuk}, A., {Moderski}, R., \& {Proga}, D. 2008, \apj, 687, 433

\bibitem[{{Janiuk} \& {Moscibrodzka}(2012)}]{2012IJMPS...8..352J}
{Janiuk}, A. \& {Moscibrodzka}, M. 2012, International Journal of Modern
  Physics Conference Series, 8, 352

\bibitem[{{Janiuk} \& {Proga}(2008)}]{janprog08}
{Janiuk}, A. \& {Proga}, D. 2008, \apj, 675, 519

\bibitem[{{Kumar} {et~al.}(2008){Kumar}, {Narayan}, \&
  {Johnson}}]{2008MNRAS.388.1729K}
{Kumar}, P., {Narayan}, R., \& {Johnson}, J.~L. 2008, \mnras, 388, 1729

\bibitem[{{Kunert-Bajraszewska} \& {Janiuk}(2011)}]{2011ApJ...736..125K}
{Kunert-Bajraszewska}, M. \& {Janiuk}, A. 2011, \apj, 736, 125

\bibitem[{{Lindner} {et~al.}(2010){Lindner}, {Milosavljevi{\'c}}, {Couch}, \&
  {Kumar}}]{2010ApJ...713..800L}
{Lindner}, C.~C., {Milosavljevi{\'c}}, M., {Couch}, S.~M., \& {Kumar}, P. 2010,
  \apj, 713, 800

\bibitem[{L{\"{o}}ffler {et~al.}(2012)L{\"{o}}ffler, Faber, Bentivegna, Bode,
  Diener, Haas, Hinder, Mundim, Ott, Schnetter, Allen, Campanelli, \&
  Laguna}]{Loffler:2011ay}
L{\"{o}}ffler, F., Faber, J., Bentivegna, E., {et~al.} 2012, Classical and
  Quantum Gravity, 29, 115001

\bibitem[{{Luminet} \& {Marck}(1985)}]{LuminetM1985}
{Luminet}, J.-P. \& {Marck}, J.-A. 1985, \mnras, 212, 57

\bibitem[{{MacFadyen} \& {Woosley}(1999)}]{1999ApJ...524..262M}
{MacFadyen}, A.~I. \& {Woosley}, S.~E. 1999, \apj, 524, 262

\bibitem[{{Mapelli} {et~al.}(2013){Mapelli}, {Zampieri}, {Ripamonti}, \&
  {Bressan}}]{2013MNRAS.429.2298M}
{Mapelli}, M., {Zampieri}, L., {Ripamonti}, E., \& {Bressan}, A. 2013, \mnras,
  429, 2298

\bibitem[{{McKinney}(2006)}]{2006MNRAS.368.1561M}
{McKinney}, J.~C. 2006, \mnras, 368, 1561

\bibitem[{{McKinney} {et~al.}(2012){McKinney}, {Tchekhovskoy}, \&
  {Blandford}}]{2012MNRAS.423.3083M}
{McKinney}, J.~C., {Tchekhovskoy}, A., \& {Blandford}, R.~D. 2012, \mnras, 423,
  3083

\bibitem[{{Merritt} {et~al.}(2004){Merritt}, {Milosavljevi{\'c}}, {Favata},
  {Hughes}, \& {Holz}}]{2004ApJ...607L...9M}
{Merritt}, D., {Milosavljevi{\'c}}, M., {Favata}, M., {Hughes}, S.~A., \&
  {Holz}, D.~E. 2004, \apjl, 607, L9

\bibitem[{Misner {et~al.}(2003)Misner, Thorne, \& Wheeler}]{gravitation}
Misner, C.~W., Thorne, K.~S., \& Wheeler, J.~A. 2003, Gravitation (W. H.
  Freeman and Company)

\bibitem[{{Piran}(2004)}]{2004RvMP...76.1143P}
{Piran}, T. 2004, Reviews of Modern Physics, 76, 1143

\bibitem[{{Podsiadlowski} {et~al.}(2004){Podsiadlowski}, {Mazzali}, {Nomoto},
  {Lazzati}, \& {Cappellaro}}]{2004ApJ...607L..17P}
{Podsiadlowski}, P., {Mazzali}, P.~A., {Nomoto}, K., {Lazzati}, D., \&
  {Cappellaro}, E. 2004, \apjl, 607, L17

\bibitem[{Schnetter {et~al.}(2004)Schnetter, Hawley, \&
  Hawke}]{Schnetter:2003rb}
Schnetter, E., Hawley, S.~H., \& Hawke, I. 2004, Class. Quantum Grav., 21, 1465

\bibitem[{{Shi} {et~al.}(2012){Shi}, {Krolik}, {Lubow}, \& {Hawley}}]{Shi:2012}
{Shi}, J.-M., {Krolik}, J.~H., {Lubow}, S.~H., \& {Hawley}, J.~F. 2012, \apj,
  749, 118

\bibitem[{Shibata \& Nakamura(1995)}]{Shibata:1995we}
Shibata, M. \& Nakamura, T. 1995, Phys. Rev. D, 52, 5428

\bibitem[{{Svensson} {et~al.}(2010){Svensson}, {Levan}, {Tanvir}, {Fruchter},
  \& {Strolger}}]{2010MNRAS.405...57S}
{Svensson}, K.~M., {Levan}, A.~J., {Tanvir}, N.~R., {Fruchter}, A.~S., \&
  {Strolger}, L.-G. 2010, \mnras, 405, 57

\bibitem[{Thornburg(2004)}]{Thornburg:2003sf}
Thornburg, J. 2004, Class. Quantum Grav., 21, 743

\bibitem[{Tichy \& Marronetti(2007)}]{tichy2007}
Tichy, W. \& Marronetti, P. 2007, Phys. Rev. D, 76, 061502

\bibitem[{{Valtonen} {et~al.}(2012){Valtonen}, {Ciprini}, \&
  {Lehto}}]{2012MNRAS.427...77V}
{Valtonen}, M.~J., {Ciprini}, S., \& {Lehto}, H.~J. 2012, \mnras, 427, 77

\bibitem[{{van Kerkwijk} {et~al.}(1992){van Kerkwijk}, {Charles}, {Geballe},
  {King}, {Miley}, {Molnar}, {van den Heuvel}, {van der Klis}, \& {van
  Paradijs}}]{Kerkwijk1992}
{van Kerkwijk}, M.~H., {Charles}, P.~A., {Geballe}, T.~R., {et~al.} 1992, \nat,
  355, 703

\bibitem[{{Vink} \& {de Koter}(2005)}]{2005A&A...442..587V}
{Vink}, J.~S. \& {de Koter}, A. 2005, \aap, 442, 587

\bibitem[{{Wellstein} \& {Langer}(1999)}]{WellsteinL1999}
{Wellstein}, S. \& {Langer}, N. 1999, \aap, 350, 148

\bibitem[{{Woosley}(1993)}]{1993ApJ...405..273W}
{Woosley}, S.~E. 1993, \apj, 405, 273

\bibitem[{{Woosley} \& {Bloom}(2006)}]{2006ARA&A..44..507W}
{Woosley}, S.~E. \& {Bloom}, J.~S. 2006, \araa, 44, 507

\bibitem[{{Woosley} \& {Heger}(2006)}]{2006ApJ...637..914W}
{Woosley}, S.~E. \& {Heger}, A. 2006, \apj, 637, 914

\bibitem[{{Woosley} \& {Weaver}(1995)}]{Woosley95}
{Woosley}, S.~E. \& {Weaver}, T.~A. 1995, \apjs, 101, 181

\bibitem[{{Xu} \& {Komossa}(2009)}]{2009ApJ...705L..20X}
{Xu}, D. \& {Komossa}, S. 2009, \apjl, 705, L20

\bibitem[{{Yoon} \& {Langer}(2005)}]{2005A&A...443..643Y}
{Yoon}, S.-C. \& {Langer}, N. 2005, \aap, 443, 643

\bibitem[{{Zdziarski} {et~al.}(2013){Zdziarski}, {Miko{\l}ajewska}, \&
  {Belczy{\'n}ski}}]{2013MNRAS.429L.104Z}
{Zdziarski}, A.~A., {Miko{\l}ajewska}, J., \& {Belczy{\'n}ski}, K. 2013,
  \mnras, 429, L104

\bibitem[{{Zhang} \& {M{\'e}sz{\'a}ros}(2004)}]{2004IJMPA..19.2385Z}
{Zhang}, B. \& {M{\'e}sz{\'a}ros}, P. 2004, International Journal of Modern
  Physics A, 19, 2385

\bibitem[{{Zhang} \& {Fryer}(2001)}]{ZhangF2001}
{Zhang}, W. \& {Fryer}, C.~L. 2001, \apj, 550, 357

\end{thebibliography}

\end{document}